# Formation of Cosmic Dust Bunnies

Lorin S. Matthews, Ryan L. Hayes, Michael S. Freed, and Truell W. Hyde, *Member, IEEE*

***Abstract*— Planetary formation is an efficient process now thought to take place on a relatively short astronomical time scale. Recent observations have shown that the dust surrounding a protostar emits more efficiently at longer wavelengths as the protoplanetary disk evolves, suggesting that the dust particles are coagulating into fluffy aggregates, "much as dust bunnies form under a bed." One poorly understood problem in this coagulation process is the manner in which micron-sized, charged grains form the fractal aggregate structures now thought to be the precursors of protoplanetary disk evolution. This study examines the characteristics of such fractal aggregates formed by the collision of spherical monomers and aggregates where the charge is distributed over the aggregate structure. The aggregates are free to rotate due to collisions and dipole-dipole electrostatic interactions. Comparisons are made for different precursor size distributions and like-charged, oppositely-charged, and neutral grains.**

## I. Introduction

DUSTY plasma, a ubiquitous component of the universe, provides a rich field for inquiry into the areas of space and laboratory physics. The coagulation of micrometer sized particles in a complex (dusty) plasma is a fundamental process that has become an increasingly important area of study not only in the context of astrophysical systems, but also in many parts of chemistry, physics, colloidal systems, and plasma processing.

The initial stage in planetesimal formation is the aggregation of dust, a process which takes place on the relatively short time scale of a few millions of years [1]. Recent astronomical evidence has shown that dust around newly formed stars emits more efficiently at longer wavelengths as the dust disk evolves which indicates that the coagulating dust forms large fluffy aggregates, "much like dust bunnies form under a bed" [2]. Several experimental and numerical studies appear to confirm this [3] – [6].

Manuscript received August 4, 2006. This work was supported in part by the National Science Foundation Grant PHY-0353558.

L. S. Matthews and T. W. Hyde are with the Center for Astrophysics, Space Physics, and Engineering Research, Baylor University, Waco, TX 76798 USA (phone: 254-710-3763; fax: 254-710-3878; e-mail: Lorin_Matthews@baylor.edu, Truell_Hyde@baylor.edu).

R. L. Hayes and M. S. Freed were 2006 REU Fellows with Baylor University, Waco, TX 76798 USA. R. L. Hayes is currently a student at Point Loma Nazarene University, San Diego, CA 92106 USA. M. S. Freed is currently a student at Northern Arizona University, Flagstaff, AZ 86011 USA.

This result is important since fractal aggregates exhibit stronger gas-grain coupling and have greater collisional cross sections due to their open nature. Although the increase in collisional cross-sectional area increases the coagulation rate, the strong gas-coupling can reduce the coagulation rate since it suppresses the relative velocities between aggregates. Thus, the physical geometry of the forming system is an essential factor in properly modeling dust coagulation.

Another factor which strongly affects the coagulation rate is the electrostatic force between dust aggregates. The dust in a proto-planetary disk is immersed in a plasma environment, produced by cosmic rays and radioactive decay [7], allowing the dust grains to become charged. This can lead to enhanced coagulation (if the grains within the population acquire opposite charges) or inhibit coagulation (if all of the grains are charged to the same potential). Recent PKE-Nefedov experiments aboard the International Space Station show evidence for the onset of "runaway growth" in which a single aggregate can collect the majority of the mass in the system; currently this is thought to be due to the charge-dipole interactions between the aggregates [8]. This phenomenon has also been seen in coagulation models of charged dust grains, in which the dipole moments of the charged aggregates and the overall dust cloud were included in the interactions [9].

In order to fully understand the evolution of planetesimals from the constituent matter in a dusty disk, it is necessary to know collision outcomes as a function of the physical parameters of the system: velocity, impact angle, aggregate material, and charge [3]. In this paper we present aggregate growth results for charged and uncharged grains and several different size distributions (monodisperse, flat polydisperse, and power law spectrum). The algorithm used explicitly tracks the charge distribution, orientation, and trajectory of each particle, allowing the dipole-dipole interactions of the charged particles to induce rotations. Only true collisions, in which the constituent particles of two colliding aggregates overlap, are detected. The algorithm also allows the history of the aggregate growth to be traced from initial monomer to the final aggregate size.

## II. Method

### A. Fractal Aggregate Characteristics

Several experimental and numerical models have been designed to study the characteristics of aggregates built from spherical monomers [4], [5]. These have shown the resulting aggregates to have fractal dimension of $d_F = 1.4 – 2.1$ and that the sticking probability is unity for monomers or clusters interacting through Brownian motion [4], [5]. The fractal dimension $d_F$ is a useful general characterization of the geometry of an aggregate, as it relates the increase in particle mass with respect to a typical radius, characteristics which in turn determine the strength of the gas-coupling. For this study, the fractal dimension of individual aggregates was approximated in the following manner: Consider a cube of side $a_0$ subdivided into many equal subboxes, each a cube of side $a << a_0$. Since $N(a) = (a_0/a)^3$ subboxes are needed to fill the original cube, the dimension of the cube, $d = 3$, is related to the number of subboxes by

$$d = \frac{\log N(a)}{\log(a_0 / a)}. \tag{1}$$

An object with fractal geometry will not occupy all of the space within the solid cube. By assigning N(a) to be the number of subboxes which contain a portion of the fractal object, the fractal dimension is then defined by

$$d_F = \frac{\log N(a)}{\log(a_0 / a)}. \tag{2}$$

In this model, aggregate characteristics are tracked employing several physical parameters including the moments of inertia, principal axes, angular velocity, charge, and center of mass velocity. The resulting representation of an aggregate consists of a data structure that contains all of the information for the aggregate and its orientation as a whole, plus a substructure containing information about the size, position, and charge of each constituent monomer relative to the aggregate's center of mass. While the aggregate characteristics are used to calculate forces and accelerations, the information on the constituent monomers is used in collision detection and calculation of the fractal dimension.

### *B. Aggregate Builder*

The algorithm used to model the collisions between particles and aggregates is similar to that described previously for N-body simulations [9, 10]. However, since the purpose of this study was to examine fractal characteristics as a function of colliding monomer's size, charge, and velocity, only potential collisions between pairs of particles were simulated.

Collisions occur in the center of mass frame of an initial seed particle, $m_1$, fixed at the origin. An incoming particle with a given mass $m_2$, charge $q_2$, and velocity $\mathbf{v}$ approaches from a randomly selected direction. The electric field of each particle is approximated using the monopole and dipole terms and the resultant acceleration calculated for each particle. The acceleration of the incoming particle is then assigned to be $\mathbf{a} = -\mathbf{a}_1 + \mathbf{a}_2$ while the acceleration of the seed particle at the origin remains zero.

The dipole moment $\mathbf{p}_i$ of a particle can also interact with the electric field (of its pair particle) $\mathbf{E}_j$, producing a torque about the original particle's center of mass

$$\mathbf{N}_i = \mathbf{p}_i \times \mathbf{E}_j. \tag{3}$$

The resulting rotational motion of the aggregate can then be calculated using Euler's equations

$$\begin{aligned} I_1\dot{\omega}_1 - \omega_2\omega_3(I_2 - I_3) &= N_1, \\ I_2\dot{\omega}_2 - \omega_3\omega_1(I_3 - I_1) &= N_2, \\ I_3\dot{\omega}_3 - \omega_1\omega_2(I_1 - I_2) &= N_3, \end{aligned} \tag{4}$$

where $I_i$, ($i$ = 1, 2, 3) are the principal moments of inertia of the aggregate and the $\omega_i$'s are its spin components as measured with respect to the body axes.

The orientation of the principal axes with respect to the stationary reference frame also changes with time, as given by the equations

$$\begin{aligned} \dot{\hat{p}}_1 &= \omega_3\hat{p}_2 - \omega_2\hat{p}_3, \\ \dot{\hat{p}}_2 &= \omega_1\hat{p}_3 - \omega_3\hat{p}_1, \\ \dot{\hat{p}}_3 &= \omega_2\hat{p}_1 - \omega_1\hat{p}_2. \end{aligned} \tag{5}$$

described in [9]. Here the $\hat{p}_i$'s are unit vectors in the direction of the principal axes.

An adaptive fifth order Runge-Kutta integrator is used to predict the updated position, velocity, spin, and orientation of each particle with a collision check performed at the end of each time step. A collision is detected only if the monomers within each aggregate physically overlap. If a collision occurs, the monomers (or aggregates) are assumed to stick together at the point of contact and the new center of mass, moments of inertia, and spin are calculated for the resultant aggregate. The total charge on the aggregate, $Q = q_1 + q_2$, is assumed to be distributed over the new aggregate structure, with a charge of

$$q_i = \frac{Qd_i}{\sum d_i} \tag{6}$$

on the i[th] monomer, where $d_i$ is the distance of the i[th] monomer from the center of mass. This distribution of charge leads to a net dipole moment for the aggregate.

The separation of electrostatic charges on the individual particles in the aggregate is not enough to disrupt the aggregate structure. The contact force between two spheres is given by $F_c = 4\pi\gamma_s r_1 r_2/(r_1 + r_2)$ [3], while the electrostatic force is $F_e = 1/4\pi\varepsilon_0\, q_1q_2/(r_1 + r_2)^2$, where $q_i = 4\pi\varepsilon_0 r_i V_i$. Assuming a value of $\gamma_s = 0.64$ for water ice [11], the ratio $F_c/F_e = 3.6 \times 10^4\, (r_1 + r_2)/V_1V_2$, where $r$ is in microns and $V$ is in volts. This ratio always exceeds unity for the range of grain sizes and potentials used in this study.

A missed collision is detected if the particles are receding from each other ($\mathbf{r} \cdot \mathbf{v} > 0$, where $\mathbf{r}$ and $\mathbf{v}$ are the position and velocity of the incoming aggregate's center of mass) and are separated by more than $10r_{max1}$, where $r_{max1}$ is the radius of a sphere centered at the seed particle's center of mass which just encloses all of the constituent monomers.

Aggregates with N ≈ 2000 monomers are formed in three generations. The first generation (up to N = 20) builds aggregates solely through the addition of incoming monomers (ballistic particle-cluster aggregation). After each collision, the aggregate's characteristics are saved to a particle library. The second generation then builds aggregates, with values of N up to 200, through collisions between particles randomly selected from the first generation library (ballistic cluster-cluster aggregation). Thus both seed and incoming aggregates can have sizes ranging from N = 2 to N = 20. Finally, the third generation builds aggregates up to N = 2000 by collisions between aggregates from the first and second generation libraries. Aggregates from the first generation library (N = 2 to N = 20) are selected 25% of the time, and aggregates from the second generation library (N = 20 to N = 200) are selected 75% of the time. Additional particles are added to the aggregate until the number of monomers exceeds the maximum number for that generation or the number of missed collisions reaches 500.

### C. *Initial Conditions*

Data was run for three different size distributions (monodisperse, flat and power law) and three different charging cases (neutral, oppositely charged, and like-charged). In the first size distribution, the particles were monodisperse with $a$ = 1.7 μm. In the second size distribution, particles ranged in size from 1 – 6 μm with a flat size distribution, while in the third size distribution, grains ranged in size from 0.5 – 10 μm assuming a power law distribution given by $n(a)\,da = a^{\gamma}\,da$, where $\gamma = -1.8$ and $n(a)\,da$ represents the number of particles with radius in the range $[a, a+da]$. For like charged grains, particles were given a potential of -1V, while for oppositely charged grains, the charge was calculated using a potential of -1 V for grains with $a < 4.0$ μm and +4 V for grains with $a \geq 4.0$ μm following Horányi and Goertz [12]. For the oppositely-charged case of the monodisperse size distribution, the overall population was selected to be charge neutral with $q_{rms} = 5000e^{-}$, to match conditions reported for the PKE-Nefedov experiment [8].

The velocity of the incoming grains was set by the minimum velocity necessary for two 0.5 μm grains charged to -1V potentials to collide when approaching from infinity, $v_{rel}$ = 0.326 m/s. The velocity of larger grains was scaled by the square root of the mass, then given a Gaussian deviation from this mean. This resulted in relative particle velocities comparable to those expected for dust grains in a protoplanetary disk with laminar flow, due to radial drift, vertical settling, and turbulence [3]. Charged dust grains have a collisional cross-section which is both velocity- and potential-dependent, with the collisional cross-section for neutral grains being larger than that for like charged grains, but smaller than that for oppositely-charged grains. To minimize CPU time spent simulating missed collisions, this was taken into account by directing the incoming particle's initial velocity towards the first particle's center of mass with a random offset $\boldsymbol{x}$. Collisions could range from direct hits, $\boldsymbol{x}$ = 0, to grazing collisions, defined as the point where the furthest extremities of the two aggregates just touch, $|\boldsymbol{x}| = (r_{max1} + r_{max2})$. The maximum offset was set to be one-half the grazing distance for neutral and oppositely-charged grains while a maximum offset of one-fourth to one-tenth the grazing distance was used in the case of like-charged grains (the lower value being used for the third generation). For comparison purposes, additional runs with the maximum offset equal to the grazing distance were also run for neutral and oppositely-charged grains with a power law size distribution.

## III. Results

The resulting fractal dimensions for aggregates formed in the nine cases described above are shown in Figs 1-3. As can be seen in Figs. 1a, 2a, and 3a, it is difficult to grow large aggregates with like-charged grains. It is interesting to note, however, that when such aggregation does occur, the fractal dimension falls much more quickly than for neutral or oppositely charged grains, indicating that only the extremities of colliding aggregates are likely to intersect. Surprisingly, parts (b) and (c) for Figs 1-3 indicate that there is little difference between the fractal dimensions of aggregates formed from the coagulation of neutral grains and that of oppositely-charged grains when the maximum offset is one-half the grazing distance.

An extended run was made to compare the results of the oppositely-charged and neutral populations assuming a power law size distribution with a maximum offset equal to the grazing distance and allowing a fourth generation of aggregates with N ≈ 10,000 to be built. Representative aggregates are shown in Figs. 4 and 5, with results for the fractal dimension shown in Fig 6. While the fractal dimension for these largest aggregates in the two cases is similar ($d_F$ = 1.6 for neutral grains and $d_F$ = 1.9 for oppositely charged grains) they clearly differ in overall structure. The neutral aggregate has a uniform distribution of monomer sizes throughout its structure, while the aggregate built from oppositely-charged particles has several very large monomers surrounded by many very small monomers. The smaller monomers fill in the "pore spaces" within the structure yielding a more compact structure which should be more resistant to crushing or fragmentation in subsequent collisions.

## IV. Conclusion and Future Work

A method for studying the formation of charged fractal aggregates has been presented. The charge is allowed to rearrange over the aggregate structure and the resulting electrostatic interactions between charged particles is approximated using the monopole and dipole moments of the charge. The particles are free to rotate due to collisions and torques induced by electrostatic interactions. Initial results were as expected: like-charged grains are less likely to coagulate. However, coagulation of oppositely-charged and neutral grains produced aggregates with very similar fractal dimensions. In studying coagulation between individual aggregates, there was no indication that a single dense aggregate would quickly increase in mass, as seen in the onset of charge gelation [8, 9]. This indicates that the onset of charge gelation is probably a property of the dipole interactions of the grains within the dust cloud, and not individual grains.

The aggregate structure is of great importance when studying the coagulation within a protoplanetary dust cloud. More open structures, such as those seen in the neutral population, will couple more strongly to the gas, reducing the relative velocities between aggregates. This will lessen the likelihood of fragmentation during collision, but at the same time will reduce the collision probability. More compact structures, such as those seen for the oppositely charged grains, will have a reduced collisional cross section, but may be more robust and less likely to fragment during collisions. Further study is necessary to determine which of these factors will have a grater influence on the coagulation rate in a protoplanetary disk.

It is difficult to put a time scale on the formation of aggregates since this model only examines potential pairs of colliding grains, and the time of each interaction that leads to a collision is the same, on average, for different charging models. However, it is instructive to look at the number of missed collisions as a function of aggregate size, as shown in Fig. 7. While the oppositely charge grains collide more efficiently in the BPCA regime (N ≤ 20), the number of missed collisions greatly exceeds that for neutral grains in the BCCA regime. This is probably due to the larger aggregates being charged to like potential as they grow, given that the initial dust population in this model is not charge neutral.

Future work will consider the time evolution of the overall charge on the fractal aggregate due to interaction with the plasma environment. This will provide a more accurate description of the total charge and the charge distribution on aggregate structures. The fractal aggregates can then be used as initial (seed) particles for an N-body simulation. Such simulations will allow the dipole moment of the dust cloud to be included in the model, which may be a driving factor in the onset of charge gelation. A comparison of coagulation timescales for the different models can then be made.

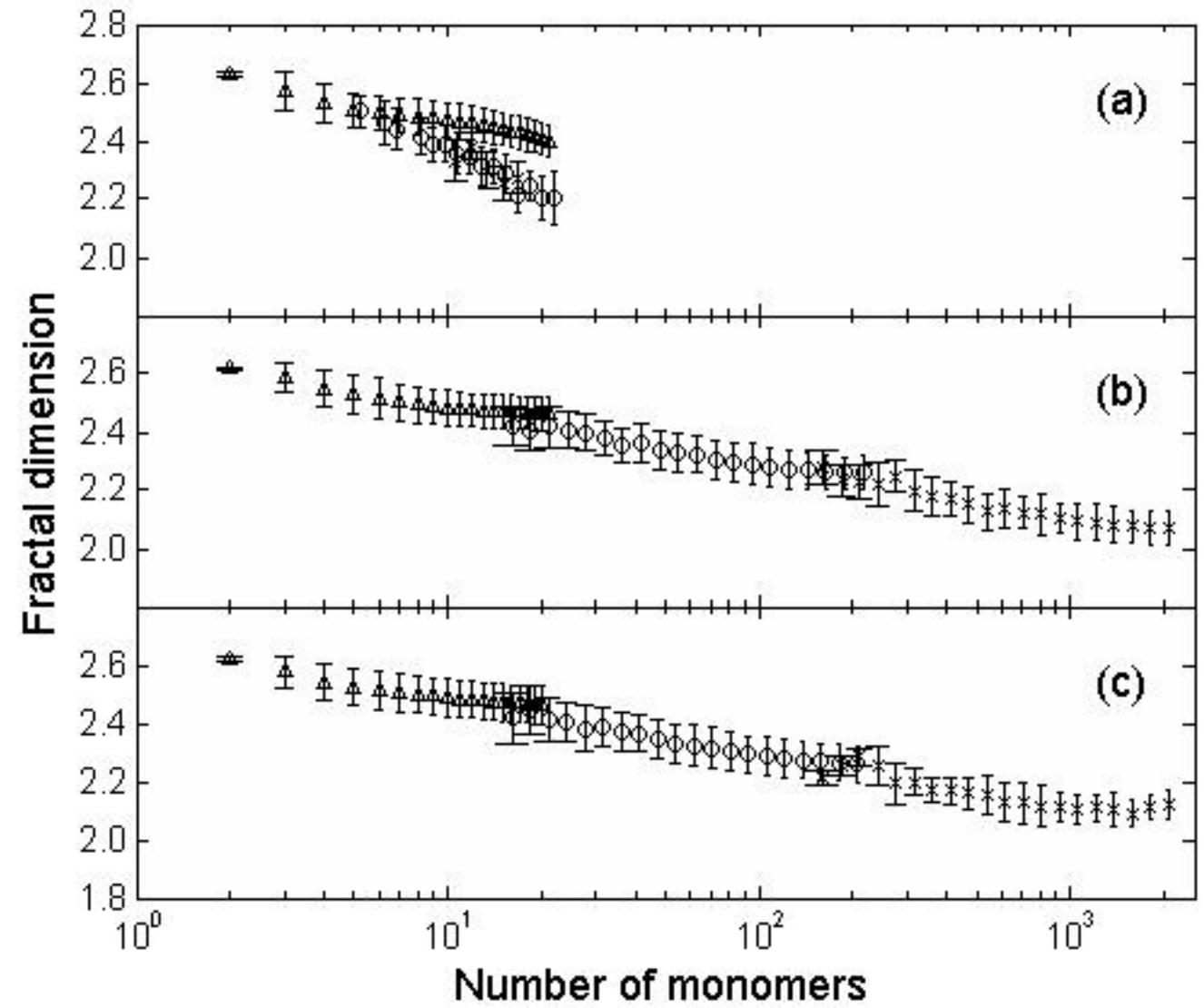

**Fig. 1.** Fractal dimensions for aggregates built from monodisperse 1.7 μm particles. The first generation of particles is designated by triangles, the second generation by circles, and the third by crosses. (a) Like-charged particles. Aggregates did not grow beyond ~30 monomers, but the aggregates that did form exhibited a much lower fractal dimension. (b) Aggregates built from neutral monomers. (c) Aggregates built from oppositely charged monomers. Interestingly, there is very little difference in the fractal dimension for these two models.

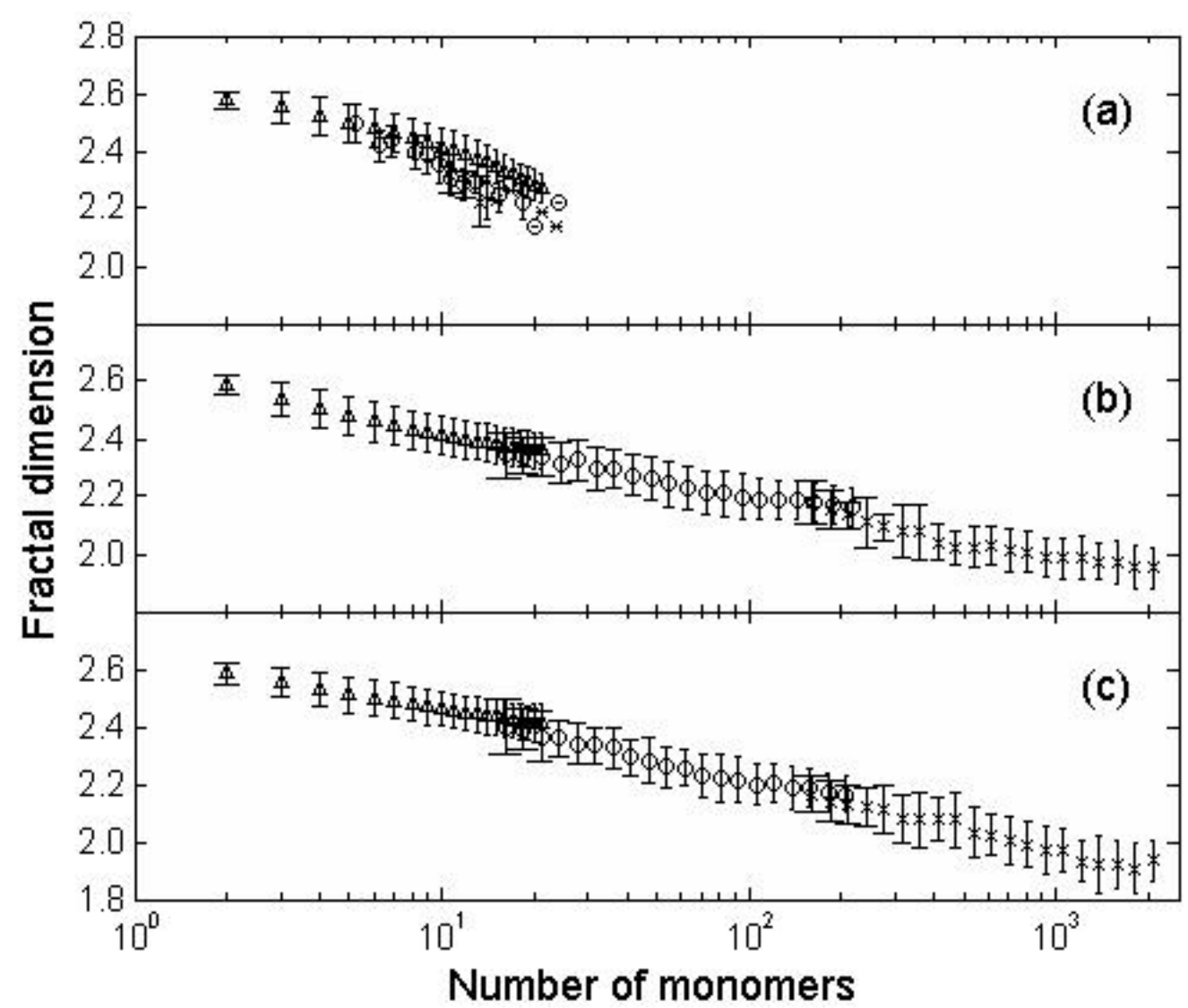

**Fig. 2.** Fractal dimensions for aggregates built from monomers with 1μm ≤ a ≤ 6 μm. The first generation of particles is designated by triangles, the second generation by circles, and the third by crosses. (a) Like-charged particles. As seen in Fig 1, the aggregates did not grow very large but the fractal dimension fell rapidly. (b) Aggregates built from neutral monomers. (c) Aggregates built from oppositely charged monomers. There is very little difference in the fractal dimension between these two models, though the average $f_D$ was slightly smaller for aggregates forming from oppositely charged grains.

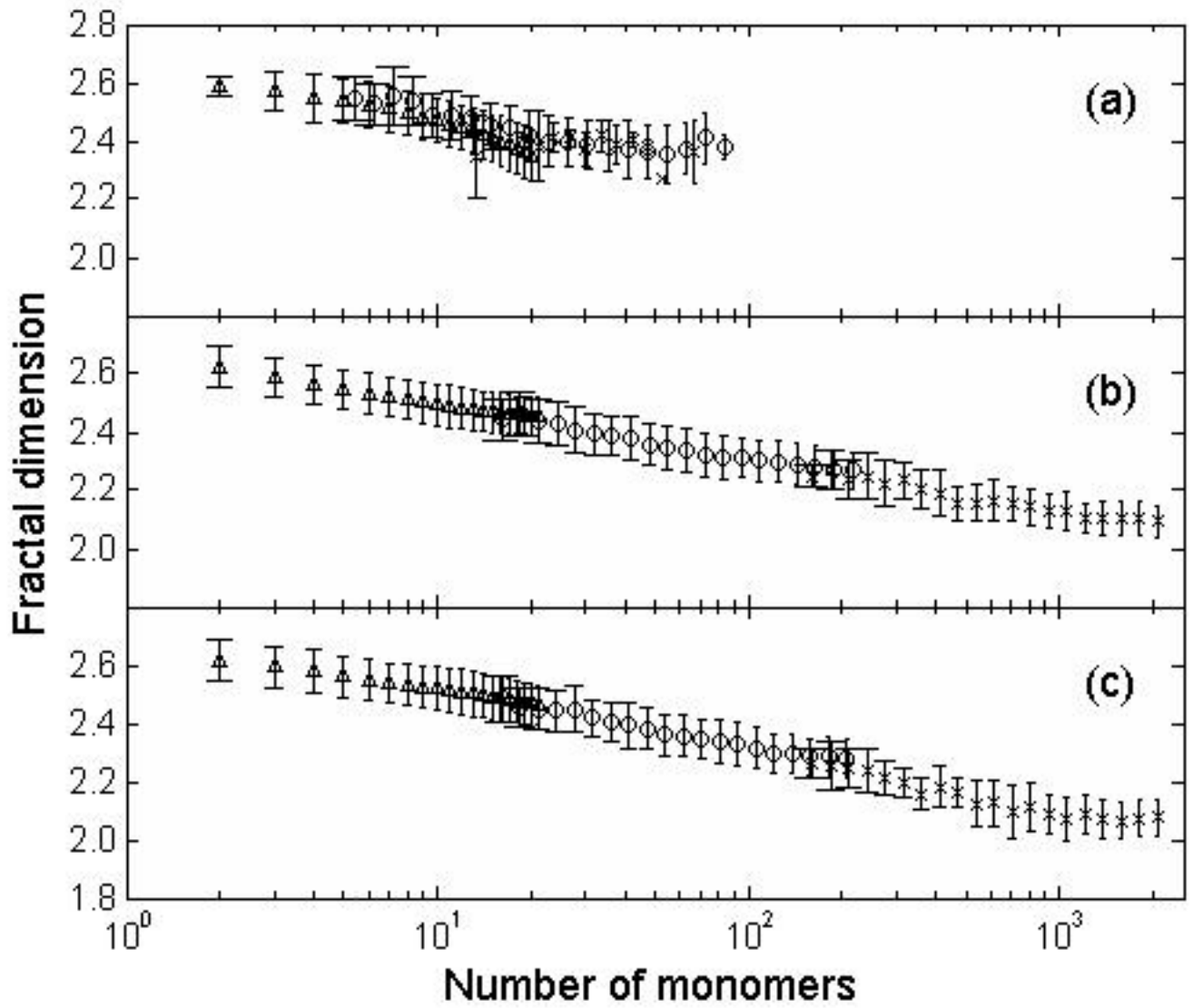

**Fig. 3.** Fractal dimensions for aggregates built from monomers assuming a power law size distribution with 0.5 μm ≤ a ≤ 10 μm. The first generation of particles is designated by triangles, the second generation by circles, and the third by crosses. (a) Like-charged particles. As seen in Fig 1, the aggregates did not grow very large but the fractal dimension fell rapidly. (b) Aggregates built from neutral monomers. (c) Aggregates built from oppositely charged monomers. There is very little difference in the fractal dimension for these two models.

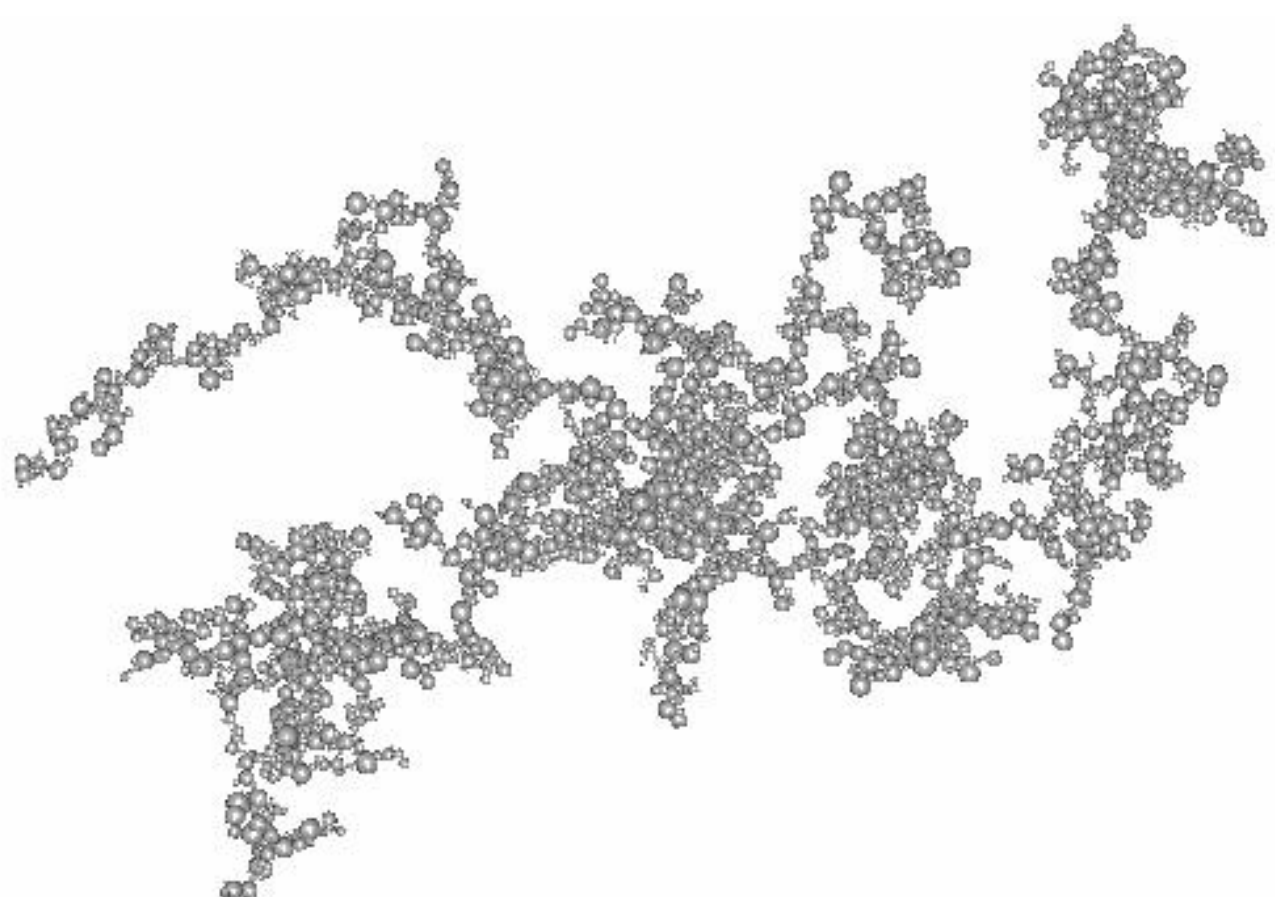

**Fig. 4.** Representative aggregate from a neutral, power law size distribution. N = 11485, $r_{max}$ = 778 μm. The size distribution of monomers throughout the aggregate structure is uniform.

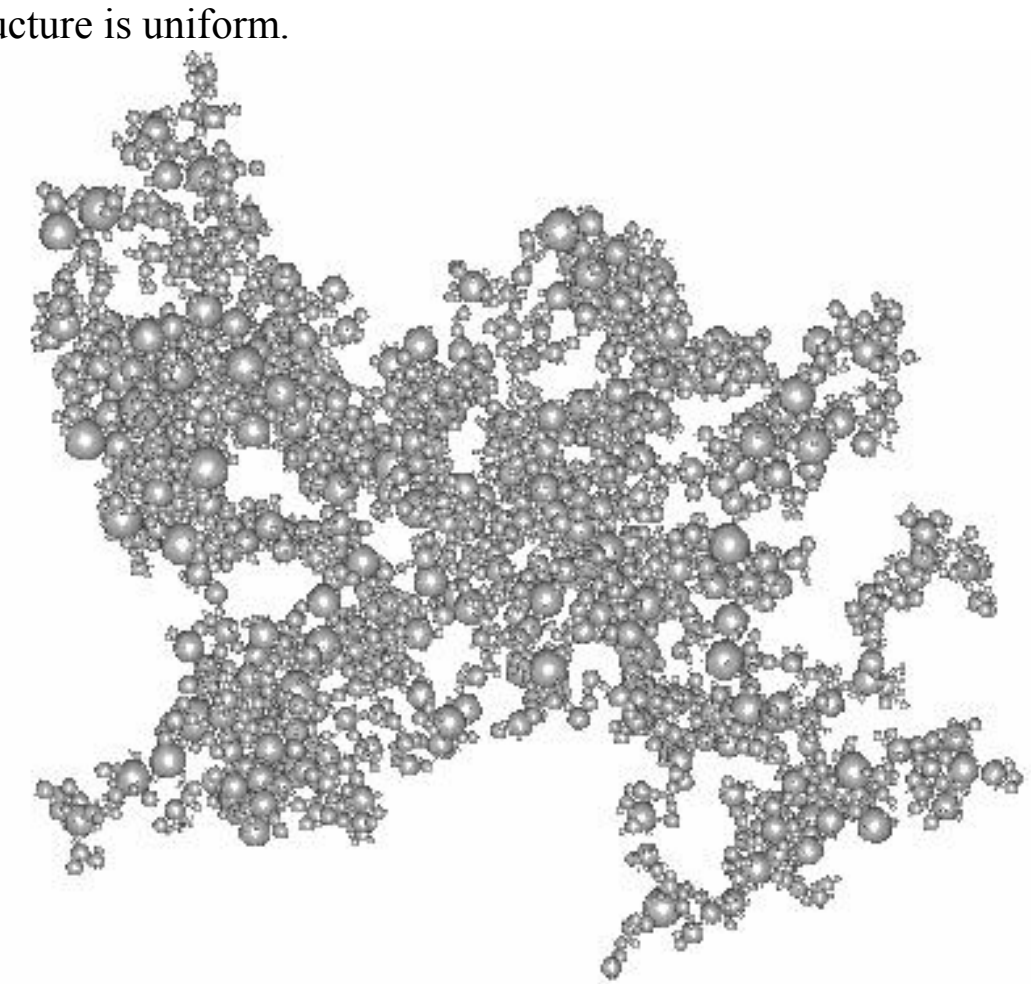

**Fig. 5.** Representative aggregate from an oppositely charged, power law size distribution. N = 11037, $r_{max}$ = 579 μm. Large monomers are surrounded by very small monomers, due to the characteristics of the initial aggregation of the oppositely charged grains.

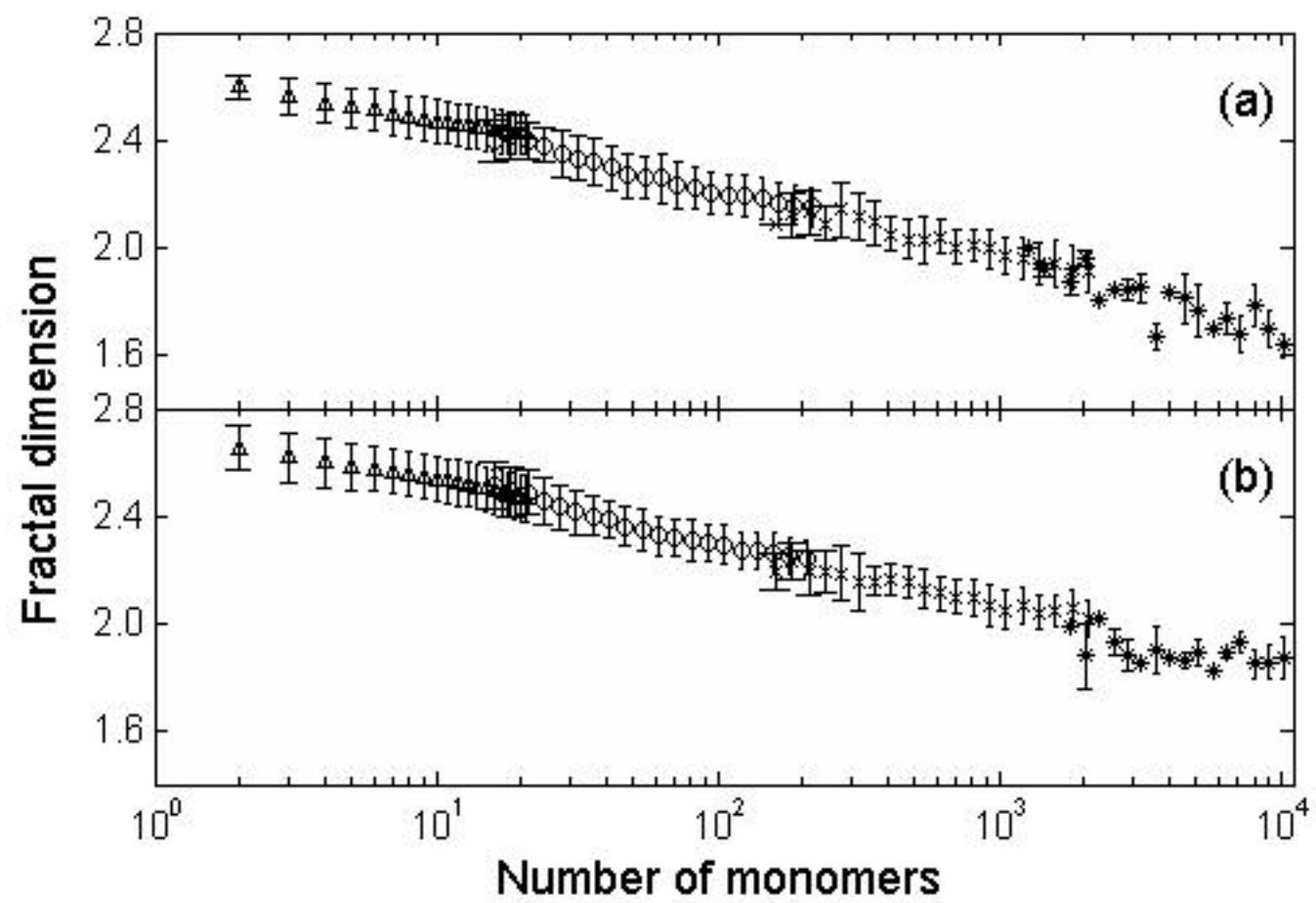

**Fig. 6.** Fractal dimensions for aggregates built from monomers with a powerlaw size distribution, 0.5 μm ≤ a ≤ 10 μm, and maximum impact parameter *x* = 1.0 (grazing collisions). The first generation of particles is designated by triangles, the second generation by circles, the third by crosses, and the fourth by asterisks. Aggregates formed from the neutral particles (a) approach a fractal dimension of 1.6, which is slightly lower than those formed from oppositely charged grains (b).

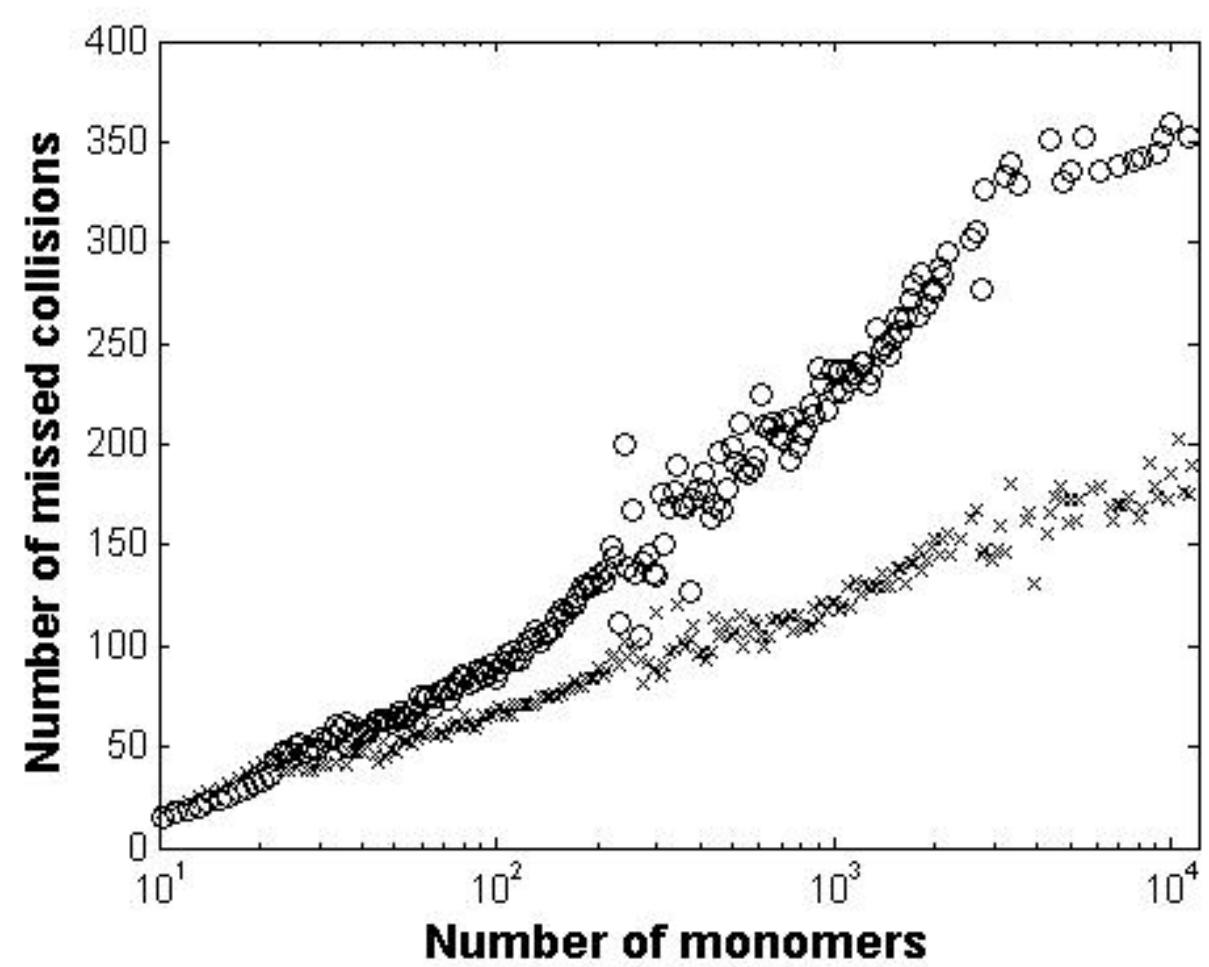

**Fig. 7.** Number of missed collisions as a function of aggregate size. Aggregates are the same as those shown in Fig. 6, with circles representing the oppositely-charged population and crosses denoting the neutral population. Oppositely-charged particles collide more efficiently in the BPCA regime (N< 20), but less efficiently in the BCCA regime as the larger clusters become charged to like potentials.

**Lorin S. Matthews** was born in Paris, TX in 1972. She received the B.S. and the Ph.D. degrees in physics from Baylor University in Waco, TX, in 1994 and 1998, respectively.

She is currently an Assistant Professor in the Physics Department at Baylor University. Previously, she worked at Raytheon Aircraft Integration Systems where she was the Lead Vibroacoustics Engineer on NASA’s SOFIA (Stratospheric Observatory for Infrared Astronomy) project.

**Truell W. Hyde** was born in Lubbock, Texas in 1956. He received the B.S. in physics and mathematics from Southern Nazarene University in l978 and the Ph.D. in theoretical physics from Baylor University in 1988.

He is currently at Baylor University where he is the Director of the Center for Astrophysics, Space Physics & Engineering Research (CASPER), a Professor of physics and the Vice Provost for Research for the University. His research interests include space physics, shock physics and waves and nonlinear phenomena in complex (dusty) plasmas.